\documentstyle[epsf,aps]{revtex}
\renewcommand{\thefootnote}{\fnsymbol{footnote}}
\begin{document}
\newcommand{\be}{\begin{eqnarray}}
\newcommand{\dlq}{\lq\lq}
\newcommand{\ee}{\end{eqnarray}}
\newcommand{\ben}{\begin{eqnarray*}}
\newcommand{\een}{\end{eqnarray*}}
\newcommand{\stackeven}[2]{{{}_{\displaystyle{#1}}\atop\displaystyle{#2}}}
\newcommand{\lsim}{\stackeven{<}{\sim}}
\newcommand{\gsim}{\stackeven{>}{\sim}}
\renewcommand{\baselinestretch}{1.0}
\newcommand{\as}{\alpha_s}
\def\eq#1{{Eq.~(\ref{#1})}}
\def\fig#1{{Fig.~\ref{#1}}}
\begin{flushright}
BNL--NT--01/12 \\
NT@UW--01--010 \\
TAUP--2682--2001\\
\end{flushright}
\vspace*{1cm} 
\setcounter{footnote}{1}
\begin{center}
{\Large\bf Instantons in the Saturation Environment}
\\[1cm]
Dmitri \ Kharzeev, $^{1}$ Yuri V.\ Kovchegov, $^{2}$ Eugene
Levin $^{3}$ \\ ~~ \\
{\it $^1$ Physics Department, Brookhaven National Laboratory} \\ 
{\it Upton, NY 11973, USA} \\ ~~ \\
{\it $^2$ Department of Physics, University of Washington, Box 351560} \\ 
{\it Seattle, WA 98195, USA } \\ ~~ \\
{\it $^3$ HEP Department, School of Physics and Astronomy } \\ 
{\it Tel Aviv University, Tel Aviv 69978, Israel } \\ ~~ \\ ~~ \\
\end{center}
\begin{abstract}
 We show that instanton calculations in QCD become theoretically well
 defined in the gluon saturation environment which suppresses large
 size instantons. The effective cutoff scale is determined by the
 inverse of the saturation scale. We concentrate on two most important
 cases: the small-$x$ tail of a gluon distribution of a high energy
 hadron or a large nucleus and the central rapidity region in a high
 energy hadronic or heavy ion collision. In the saturation regime the
 gluon density in a single large ultrarelativistic nucleus is high and
 gluonic fields are given by the classical solutions of the equations
 of motion. We show that these strong classical fields do not affect
 the density of instantons in the nuclear wave function compared to
 the instanton density in the vacuum. A classical solution with
 non-trivial topological charge is found for the gluon field of a
 single nucleus at the lowest order in the instanton perturbation
 theory. In the case of ultrarelativistic heavy ion collisions a
 strong classical gluonic field is produced in the central rapidity
 region. We demonstrate that this field introduces a suppression
 factor of $\exp \left\{ - c \, \rho^4 Q_s^4 \, / \, [ 8 \, \as^2 \,
 N_c \, (Q_s \tau)^2 ] \right\}$ in the instanton size distribution,
 where $Q_s$ is the saturation scale of both (identical) nuclei,
 $\tau$ is the proper time and $c \approx 1$ is the gluon liberation
 coefficient. This factor suggests that gluonic saturation effects at
 the early stages of nuclear collisions regulate the instanton size
 distribution in the infrared region and make the instanton density
 finite by suppressing large size instantons.
\end{abstract}
\renewcommand{\thefootnote}{\arabic{footnote}}
\setcounter{footnote}{0}

\section{Introduction}

At very high energies corresponding to extremely small values of
Bjorken $x$ variable the density of partons in the hadronic or nuclear
light cone wave function can get very large leading to the effect of
{\it saturation} of gluon and quark distributions \cite{glrmq}. The
large density of partons in the transverse plane produces strong
gluonic fields $A_\mu \, \sim \, 1/g$ which leads to a number of
interesting non-linear phenomena. The transition to the saturation
regime in the $(x,Q^2)$ plane is characterized by the saturation scale
$Q_s (x)$. It has been shown in \cite{LR,Mue,mv,agl} that for a large
nucleus the saturation scale grows with the nuclear atomic number as
$Q_s^2 \, \sim \, A^{1/3}$, and thus for a heavy ion it can get large
($Q_s^2 \, \gg \Lambda_{QCD}^2$) making the strong coupling constant
small $\as (Q_s) \, \ll \, 1$. This allowed McLerran and Venugopalan
\cite{mv} to argue that gluonic fields in the saturation region are
given by the solution of the classical Yang--Mills equations of
motion in the presence of the source given by the ultrarelativistic
nucleus on the light cone. The solution of the classical Yang--Mills
equations for a large nucleus (non-Abelian Weizs\"{a}cker-Williams
field) has been found in \cite{yuri,jklw}. It has been shown
\cite{LR,jklw,KM} that saturation regularizes the power-law
divergence of the classical unintegrated gluon distribution function
in the infrared region. The unintegrated gluon distribution at the
two-gluon level is proportional to $1/k_\perp^2$ and diverges for
small $k_\perp$. The classical field resums all multiple rescattering
effects, which corresponds to summing up powers of $Q_s/k_\perp$. The
resulting unintegrated gluon distribution given by the non-Abelian
Weizs\"{a}cker-Williams field of a large nucleus is proportional to
$\ln Q_s/k_\perp$ in the small transverse momentum region
\cite{jklw,KM}. Thus the infrared singularity becomes only
logarithmic and integrable. As we go towards smaller values of $x$
quantum corrections to the classical field become important. The
summation of leading logarithmic ($\ln 1/x$) corrections to the
classical multiple rescattering picture led to a non-linear evolution
equation for the total scattering cross section of a color dipole on a
hadron or nucleus which was derived independently by one of the
authors in \cite{yurieq} using Mueller's dipole model of \cite{dip}
and by Balitsky in \cite{bal} employing the high energy effective
lagrangian technique. Similar equation has emerged recently out of
renormalization group approach \cite{JKLW,cons}. The solution of this
non-linear evolution equation should give the behavior of the hadronic
cross sections at very high energies and specify the dependence of
$Q_s$ on $x$. The solution has been found by approximate analytical
methods \cite{yurieq,lt} and by numerical simulations \cite{braun,lub}
yielding us with the saturation scale which is a growing function of
energy, $Q_s \, \sim \, 1/x^\delta$, where $\delta$ is close to the
value of the BFKL pomeron intercept \cite{BFKL}. Thus at very high
energies the saturation scale can get large even for a hadron making
the small-$x$ tail of the gluon distribution in hadrons similar to the
small-$x$ tail of the gluon distribution in large nuclei.

With a large fraction of high energy scattering data being attributed
to some non-perturbative QCD phenomena, instantons
\cite{bpst,th,cdg,jr} might play an important role in high energy
scattering processes \cite{bj}. There have been developed several
techniques of calculating the instanton-mediated scattering amplitudes
including the effective instanton lagrangian approach \cite{cdg1,svz}
and the valley method \cite{val}. Instanton-induced effects could be
important to deep inelastic scattering processes and other high energy
processes as was shown in \cite{bb,rs,sz}. Several years ago there has
been developed a vigorous activity studying the possibility of baryon
number violating effects due to instantons in electroweak theory
\cite{bnv}. Recently the authors suggested that instanton-induced
interactions could play a key role in the dynamics of soft pomerons in
hadronic scattering cross sections and in multi-particle production
\cite{kkl} (see also \cite{nsz}). 

The role of instantons in heavy ion collisions has not been studied in
such detail. It has been recently suggested \cite{ce} that the
instanton-induced particle production can account for a significant
fraction of the total charged particle multiplicity in the heavy ion
collisions at RHIC \cite{pho}. Since it has been argued that the
small-$x$ gluons are most important for the mid-rapidity particle
production in heavy ion collisions \cite{claa,md,kv,yuri1} it is
natural to investigate whether instanton-induced effects are important
for these small-$x$ gluons, which is our main goal here.

The qualitative picture of what one might expect to happen to
instantons in the saturation environment is as follows. Let us
consider a gluon propagating through the strong background gluonic
field created either by a single large nucleus or in a collision of
two nuclei. The gluon would undergo multiple rescatterings in the
field which would have the physical effect of generating some non-zero
effective mass for the gluon, which would be roughly proportional to
$Q_s$ (see, for instance, \cite{Muee}). Equivalently we can say that
due to multiple rescatterings the gluons are screened on the
transverse distances inversely proportional to the effective mass,
i.e., on the distances of the order of $1/Q_s$. Here we can draw an
analogy to the case of QCD at high finite temperature $T$. There the
Debye mass of the gluons in quark-gluon plasma $m_D \, \sim \, g T$
introduces screening on the distances of the order of $1/m_D \, \sim
\, 1/g T $ which in turn suppresses large size instantons by the
factor of \cite{gpy}
\be\label{fint}
e^{- const \, \rho^2 \, m_D^2 } \, \sim \, e^{ - const' \, \rho^2 \, T^2 }.
\ee
In these approximate estimates we are not keeping track of the factors
of $g$, $N_C$ and $N_f$, putting them to be some constants. Thus
arguing that the saturation scale plays the role of Debye screening
mass for small-$x$ gluons $m_D \, \sim \, Q_s$ \cite{bmss} we might
expect that the instanton size distribution in the saturation
background would also obtain a suppression factor of
\be\label{goal}
e^{- const \, \rho^2 \, Q_s^2}
\ee
for large instantons. This would imply that instantons with sizes
$\rho \, \gsim \, \rho_0 \sim 1/Q_s$ are exponentially
suppressed and only small instantons with sizes $\rho \, \lsim \,
\rho_0$ can contribute to the scattering processes. This would justify 
the perturbative calculation we are about to perform below, since the
strong coupling constant for small size instantons is small $\as (\rho
< 1/Q_s) \, \ll \, 1$.

Suppression of large size instantons could be anticipated based on the
similar arguments related to the philosophy introduced in \cite{mv}:
the typical size of the nuclear color charge fluctuations in the
transverse plane is of the order of $1/Q_s$. Thus instantons of larger
size would just be washed out by the color charge fluctuations in the
ultrarelativistic nucleus.

An analogy can also be drawn with QCD at finite density and zero
temperature. There the instantons are again suppressed by an
exponential factor of \cite{finm}
\be\label{finm}
e^{- const \, \rho^2 \, \mu^2}.
\ee
The suppression is due to the following qualitative picture: large
size instantons tend to produce quarks with momenta $k \, \sim \,
1/\rho \, \lsim \, \mu$ and this process is suppressed at large $\mu$
due to Pauli blocking since all the quark energy levels with momenta
$k \, \le \, \mu$ are occupied \footnote{Yu. K. would like to thank
Larry Yaffe for providing this argument.}. Instantons in the
saturation environment produce gluons, which are of course bosons and
do not have any Pauli blocking effects. The role of Pauli blocking for
gluons is played by the distribution of gluons in the saturation wave
function of a single nucleus which is very similar (at the qualitative
level) to the Fermi-Dirac distribution. In the case of nuclear
collision the transverse momentum spectrum of produced gluons also
exhibits saturation behavior and levels off in the infrared resembling
Fermi-Dirac distribution \cite{yuri1}. Due to multiple rescatterings
the gluons with lower transverse momenta get pushed towards higher
transverse momenta of the order of $k_\perp \, \sim \, Q_s$. Thus
there are very few gluons left with small $k_\perp$ and the large size
instanton-like solutions which would tend to produce those should be
suppressed. Associating the width of the nuclear gluon distribution
with the chemical potential of the Fermi-Dirac distribution $Q_s \,
\sim \, \mu$ we would again arrive at \eq{goal}.

The intuitive arguments presented above suggest that the interaction
with gluons in the saturation environment will exponentially suppress
the instantons with large size (larger than $1/Q_s(x)$).  Such
suppression would result in an infrared safe and, therefore, well
defined theoretically instanton calculus in the saturation medium. To
show that this is indeed the case is the main objective of this paper.

The paper is organized as follows. In Sect. II we present the general
techniques that we are going to use. We will briefly review the
solution of the problem of small instantons in a slowly varying
background field using the effective instanton lagrangian approach
\cite{cdg1,svz}. We show that the background field affects the
instanton size distribution only if it has a non-zero value of
$G_{\mu\nu}^{a2}$ (or $G_{\mu\nu}^{a} \tilde{G}_{\mu\nu}^{a}$) at the
space-time point of interest \cite{cdg1,svz}. 

We will proceed in Sect. III by addressing the issue of instanton size
distribution in the small-$x$ tail of the gluon distribution in a
single nucleus or hadron.  Throughout most of this paper we will
consider the classical picture of saturation. We are interested in the
small-$x$ tail of the distribution with $x$ small enough for multiple
rescattering effects to be become important
\cite{mv,yuri,jklw,KM}. For this we require the coherence length of a
small-$x$ gluon to be comparable to nuclear size $l_c \, \approx \,
1/2 m x \, \gsim \, 2 R$, which yields us with $x
\, \lsim \, 1/ 4 m R$ \cite{LR,BM,fs}. At the same time the value of $x$ should 
{\it not} be too small since we do not want the quantum corrections
to start playing an important role. Thus we want $\as \, \ln 1/x \,
\lsim 1$ which leads to $x \, \gsim \, \exp ( - 1/\as)$. In
Sect. IIIA we demonstrate that the classical gluon field of a single
hadron or nucleus has zero value of $G_{\mu\nu}^{a2}$ and does not
affect the instanton size distribution, introducing no enhancement or
suppression of instanton effects. We proceed in Sect. IIIB by
calculating the one loop leading logarithmic ($\ln 1/x$) correction to
the classical field of the nucleus. For this correction to become
important we have to relax the $\as \, \ln 1/x \, \lsim 1$
condition. We observe that even after inclusion of this correction the
field strength squared $G_{\mu\nu}^{a2}$ of the gluon field still
remains zero and does not affect instanton distribution. We relate
this observation to the representation of quantum evolution as a
series of classical emissions as advocated in \cite{dip,JKLW} and
based on that argue that our result is true to all orders in evolution
resumming powers of $\as \, \ln 1/x$ \cite{yurieq,bal}.

The classical field of a large nucleus found in \cite{yuri,jklw} was a
non-topological solution, in the sense that it did not create a
transition from a region of one topological charge to a region of a
different topological charge. In this paper we would like to address
the issue whether there exists a topological solution of the classical
Yang-Mills equations of motion in the presence of external source
given by the large nucleus. In Sect. III we show that this solution,
if exists, is just as probable as the usual ``non-topological''
solution of \cite{yuri,jklw}. Unfortunately we were unable to
construct an exact instanton-like solution for this problem. Instead
in Sect. IIIB we are going to consider a QCD instanton perturbatively
interacting with the classical field of the nucleus \cite{yuri,jklw}
and producing a combined classical field of the instanton--nucleus
configuration which would now connect regions of different topological
charge. This field squared contributes to the one loop correction to
$G_{\mu\nu}^{a2}$, and, if viewed as a field of the nucleus probed by
a point-like instanton is not a classical field anymore. However the
field itself is a classical field of the instanton--nucleus
configuration. This classicality is due to the fact that we will be
performing the calculation resumming all powers of the parameter
$\as^2 \, A^{1/3}$ (or, equivalently, $Q_s^2 / k_\perp^2$) and
neglecting higher order corrections in $\as$ to it \cite{yuri,KM}. The
equivalence between the classical field techniques and this
resummation has been discussed in
\cite{mv,yuri,jklw,KM,claa,md,kv,yuri1} and references therein.

In Sect. IV we will consider the case of hadronic or nuclear
collisions. As has been argued in \cite{claa} the dominant particle
production mechanism for central rapidity gluons could be due to the
strong classical gluon field produced by the colliding hadrons or
nuclei. The field is again characterized by the saturation scale of
the colliding nuclei $Q_s$. This gluonic field has been calculated at
the lowest order in perturbation theory in \cite{claa,md} and analyzed
numerically in \cite{kv}. Recent progress in calculating the produced
particle spectrum due to the classical field was made in
\cite{yuri1}. The classical field produced by colliding nuclei has a
non-vanishing value of $G_{\mu\nu}^{a2}$ and therefore influences the
instanton size distribution. By a direct calculation using the lowest
order classical field \cite{claa,md} we demonstrate in Sect. IV that
the effect of this classical fields would be to suppress large size
instantons by a suppression factor of
\be\label{aasup}
\exp \left( - \frac{c \, \rho^4 Q_s^4 }{ 8
\, \as^2 \, N_c \, (Q_s \tau_0)^2 } \right), 
\ee
with the proper time of the instanton position $\tau_0 = \sqrt{2
x_{0+} x_{0-}}$ and $c$ the gluon liberation coefficient. $x_0$ is the
space-time point where we measure the instanton density. \eq{aasup}
shows that at large proper time ($\tau \rightarrow \infty$), long
after the collision, the lagrangian density $G_{\mu\nu}^{a2}$ at each
particular point gets small and large instantons will not be
suppressed anymore. The amount of suppression depends on the gluon
liberation coefficient, which was estimated numerically to be $c
= 1.29 \pm 0.09$ \cite{kv} and analytically to be $c \approx 2 \ln 2$
\cite{yuri1}, while RHIC data \cite{pho} suggests $c = 1.23 \pm 0.20$
\cite{KN}. We will also propose that the instanton suppression of 
\eq{aasup} at extremely high energies might lead to suppression and, 
therefore, unitarization of the soft pomeron of \cite{kkl}. We will
conclude the paper by estimating that the saturation effects reduce
the instanton density in the central rapidity region at RHIC by three
orders of magnitude compared to the instanton density in vacuum as
extracted from lattice data
\cite{rs}.

\section{Instantons in Background Field}

The problem of small instantons in a slowly varying background field
was first addressed in \cite{cdg1,svz} and was resolved by introducing
the effective instanton lagrangian $L^{I(\overline{I})}_{eff}
(x)$. The complete field of a single instanton solution could be
reconstructed by perturbatively resumming the powers of the effective
instanton lagrangian which corresponds to perturbation theory in
powers of the instanton size parameter $\rho^2$. In our case here the
background field arises due to the strong source current
$J_\mu^a$. The current will be due to a single nucleus (Sect. III) and
two colliding nuclei (Sect. IV). Perturbative resummation of powers of
the source current term translates itself into resummation of the
powers of the classical field parameter $\as^2 A^{1/3}$
\cite{mv,yuri}. Thus the problem of instantons in the background 
classical gluon field is described by the effective action in
Minkowski space
\be\label{qcdi}
S_{eff} \, = \, \int d^4 x \left( - \frac{1}{4} G^a_{\mu\nu} (x)
G^a_{\mu\nu} (x) \, + \, L^I_{eff} (x) \, + \, L^{\overline{I}}_{eff}
(x) \, + \, J_\mu^a \, A_\mu^a (x) \right).
\ee
In \eq{qcdi} the point-like instanton vertices are given by the
instanton-induced effective lagrangian \cite{cdg1,svz,abc}
\be\label{effl}
L_{eff}^I (x_0) \, = \, \int d \rho \, n_0 (\rho) \, dR \, \exp \left(
- \frac{2 \pi^2}{g} \, \rho^2 \, \overline{\eta}^M_{a\mu\nu} \, R^{a
a'} \, G^{a'}_{\mu\nu} (x_0) \right)
\ee
in which $n_0 (\rho)$ is the instanton size distribution function in
vacuum given by \cite{size,ss}
\be\label{isize}
n_0 (\rho) \, = \, \frac{0.466 e^{-1.679 N_c}}{(N_c - 1)! (N_c - 2)!} \, 
\frac{1}{\rho^5} \, \left( \frac{2 \pi}{\as (\rho)}
\right)^{2 N_c} \, e^{- \frac{2 \pi}{\as (\rho)}} ,
\ee
where $b \, = \, (11/3) \, N_c \, - \, (2/3) \, N_f$. In \eq{effl}
$\overline{\eta}^M_{a\mu\nu}$ is the 't Hooft symbol in Minkowski
space defined in terms of the usual 't Hooft symbol in euclidean space
by \cite{abc}
\be
\overline{\eta}^M_{a\mu\nu} \, = \, \left\{ \begin{array}{c} 
\overline{\eta}_{a\mu\nu}, \hspace*{1cm} \mu, \, \nu \, = \, 1, 2, 3 
\\ \\ \\   i \overline{\eta}_{a4\nu}, \hspace*{.5cm} \mu = 0, \, \nu \, 
= \, 1, 2, 3. \end{array} \right.
\ee
$x_0$ in \eq{effl} is the position of the instanton and $R^{a a'}$ is
the matrix of rotations in the color space with $dR$ denoting the
averaging over instanton color orientations. To obtain the effective
lagrangian for anti-instantons from \eq{effl} one has to change
$\overline{\eta}^M_{a\mu\nu}$ into $\eta^M_{a\mu\nu}$ in it
($\eta^M_{a\mu\nu} \, = \, (\overline{\eta}^M_{a\mu\nu})^\ast$).

The classical current of a single large ultrarelativistic nucleus in
the effective action of \eq{qcdi} is given in McLerran-Venugopalan
model by \cite{mv,yuri,jklw}
\be\label{cur}
J^a_\mu \, = \, \delta_{\mu +} \, \delta(x_-) \, \rho^a (x_\perp),
\ee
where $\rho^a (x_\perp)$ is the two-dimensional color charge density
of the nucleus. In the calculations of diagrams below we will be using
the explicit model of the nucleus as consisting of independent
nucleons which is justified at high energies and is equivalent to the
description of the nucleus in terms of the current of \eq{cur}
\cite{yuri,KM}. In the case of two colliding nuclei the source current will be
\be
J^a_\mu \, = \, \delta_{\mu +} \, \delta(x_-) \, \rho_1^a (x_\perp) \, +
\, \delta_{\mu -} \, \delta(x_+) \, \rho_2^a (x_\perp),
\ee
with $\rho_1^a (x_\perp)$ and $\rho_2^a (x_\perp)$ the color charge
densities of the colliding nuclei. 

The action of \eq{qcdi} solves the problem of topologically
non-trivial classical fields of nuclei at the conceptual level. It
allows one to construct a perturbative series in the powers of
$\rho^2$ and $\as^2 A^{1/3}$ which, with proper regularization of
singularities, should sum up to yield us the classical field of a
nucleus or nuclei with non-zero topological charge. In Sect. IIIB we
will construct an example of such field in the single instanton sector
at the lowest order in these parameters.

Following \cite{cdg1,svz} we may write the first correction in $\rho$
to the instanton size distribution generated by the background field
of the nucleus or nuclei as
\ben
n_{sat} (\rho) \, = \, (A| L^I_{eff} |A) \, = 
\een
\be\label{effn}
= \, n_0 (\rho) \, \left( 1 \, + \, \frac{\pi^3 \rho^4}{\as (N_c^2 -
1)} \, (A|\ G^{a}_{\mu\nu} (x_0) \, G^{a}_{\mu\nu} (x_0) - \,
G^{a}_{\mu\nu} (x_0) \tilde{G}^{a}_{\mu\nu} (x_0) \ |A) \, +
\,\mbox{higher orders in} \, \rho^4 \right)
\ee
where $G^{a}_{\mu\nu} (x_0)$ is the field strength of the slowly
varying background field taken at the position of the instanton and
$\tilde{G}^{a}_{\mu\nu} = (1/2) \epsilon_{\mu\nu\rho\sigma}
G^{a}_{\rho\sigma}$ is the dual field strength. $(A|
\ldots |A)$ denotes averaging in the nuclear wave function(s)
\cite{mv,yuri}. \eq{effn} could be obtained by expanding the effective 
lagrangian $L^I_{eff}$ in the powers of $\rho^2$ and averaging the
resulting terms in the nuclear wave function and over instanton
orientations \cite{cdg1,svz,abc}. As was shown in \cite{yuri} for the
classical field of a large nucleus at the leading powers in $A$ the
higher order correlators factorize (Gaussian averaging) yielding the
property which was assumed for the vacuum fields in \cite{svz}
\ben
(A| \ \overline{\eta}^M_{a_1\mu_1\nu_1} \, R^{a_1 a'_1} \,
G^{a'_1}_{\mu_1\nu_1} \, \ldots \,
\overline{\eta}^M_{a_{2k}\mu_{2k}\nu_{2k}} \, R^{a_{2k} a'_{2k}} \,
G^{a'_{2k}}_{\mu_{2k}\nu_{2k}} \ |A) \, =
\een
\be\label{fact}
= \, (2
k - 1)!! \, \left[ (A| \ \overline{\eta}^M_{a\mu\nu} \, R^{a a'}
\, G^{a'}_{\mu\nu} \, \overline{\eta}^M_{b\alpha\beta} \, R^{b b'}
\, G^{b'}_{\alpha\beta} \ |A) \right]^k.
\ee
Using the color neutrality of a nucleus we write
\be\label{neu}
(A| \ G^{a'}_{\mu\nu} \, G^{b'}_{\alpha\beta} \ |A) \, = \,
\frac{\delta^{a'b'}}{N_c^2 - 1} \, (A| \ G^{a}_{\mu\nu} \, 
G^{a}_{\alpha\beta} \ |A).
\ee
Employing \eq{neu} in \eq{fact} together with the orthogonality of the
color rotation matrices $R^{a a'} \, R^{b a'} = \delta^{ab}$ and
performing some simple algebra of 't Hooft symbols one can show that
the effect of higher order terms in $\rho^4$ in \eq{effn} is just to
exponentiate the lowest order term leading to (cf. \cite{cdg1,svz})
\be\label{effnf}
n_{sat} (\rho) \, = \, n_0 (\rho) \, \exp \left[ \frac{\pi^3
\rho^4}{\as (N_c^2 - 1)} \, (A|\ G^{a}_{\mu\nu} (x_0) \,
G^{a}_{\mu\nu} (x_0) - \, G^{a}_{\mu\nu} (x_0) \tilde{G}^{a}_{\mu\nu}
(x_0) \ |A) \right].
\ee
Thus in order to find the effect of a particular background field on
the instanton size distribution all one has to do is to calculate the
matrix element in the power of the exponent in \eq{effnf}. The
effective instanton lagrangian approach is strictly valid only for
instantons of the size much smaller than the typical variation of the
external field \cite{svz} though it usually works for larger
instantons too. In the case of saturation background there is only one
dimensional parameter characterizing the external field and it is the
saturation scale $Q_s$. Thus one might argue that the matrix element
of $G^{a2}_{\mu\nu}$ is proportional to $Q_s^4$ and therefore
\eq{effnf} resums all powers of the parameter $\rho^4 Q_s^4$ and is
strictly speaking valid only when
\be\label{apprx}
\rho^4 \, Q_s^4 \, \ll \, 1.
\ee
This is the approximation that we are going to employ throughout the
paper: the instantons we consider are much smaller than $1/Q_s$. This
assumption also keeps $\as (\rho)$ small allowing us to use
perturbation theory.

\eq{naa} includes only the dipole interaction term between the
instanton and the background field. Higher order multipole
interactions in general should also be included. However, since these
interactions would correspond to subleading in $A$ correction to
factorization of \eq{fact} the multipole terms would also be
subleading in $A$. They would have less powers of $A^{1/3}$ per power
of $\rho^4$ than the leading (dipole) term and could be neglected.

\section{Single Nucleus Case}

\subsection{Lowest Order Diagrams}

Let us start constructing the modified instanton size distribution in
the external field of a single nucleus along with the field of the
instanton-nucleus configuration using the action of \eq{qcdi}. We will
consider a single instanton interacting with the classical field of a
large ultrarelativistic nucleus. At the lowest orders in $\rho^2$ and
$g$ the field of the I-nucleus configuration would be given simply by
the sum of the fields of the instanton and the nucleus, similarly to
the sum ansatz for the I--$\overline{\mbox{I}}$ configurations
\cite{dp}. Since we are interested in the effect of the nuclear gluon
field on the instanton size distribution it is more convenient to
start analyzing diagrams for the interaction of the instanton with the
field directly, without first deriving the field. This is equivalent
to calculating the matrix element in the exponent of \eq{effnf}. We
will also concentrate for now on the case of a single rescattering in
the nucleus, which parametrically corresponds to the case of $\as^2
A^{1/3} \, \lsim \, 1$. We will generalize our results to the $\as^2
A^{1/3} \, \sim \, 1$ case after we obtain the lowest order
expression.

\begin{figure}
\begin{center}
\epsfxsize=12cm
\leavevmode
\hbox{ \epsffile{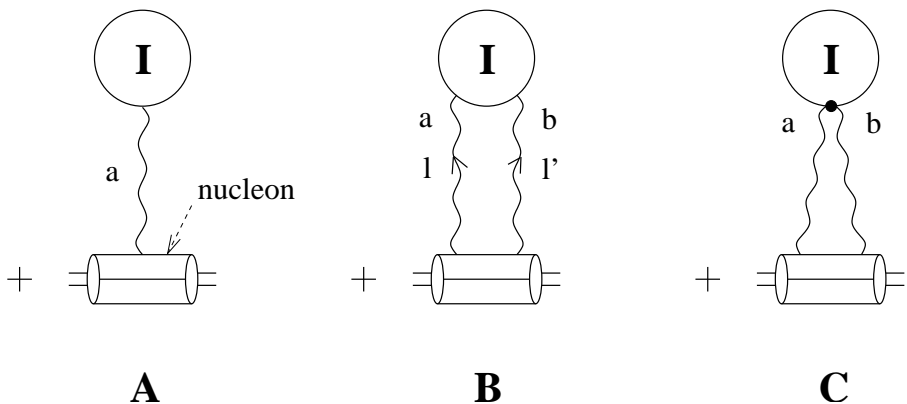}}
\end{center}
\caption{Lowest order diagrams which might contribute to the instanton 
action in the field of a nucleus. Only one nucleon participates in the
interaction here. After a simple calculation (see text) one can see
that the contributions of these diagrams are zero.  }
\label{lo}
\end{figure}

All the graphs that we are going to analyze will be calculated in
$\partial_\mu A_\mu \, = \, 0$ light cone gauge. The lowest order
diagram which might contribute to the action of the I-nucleus
configuration is the one gluon exchange diagram depicted in
\fig{lo}A. There an instanton interacts with one nucleon in the
nucleus which is moving ultrarelativistically with the large momentum
in the light cone ``$+$'' direction. To obtain the average value of
the action we have to average this diagram (as well as all others) over 
the nuclear wave function \cite{mv,yuri,jklw,KM}. The procedure
includes averaging over all possible positions of nucleons in nuclei
and quarks in nucleons, as well as averaging in the color spaces of
each particular nucleon, which makes sure that the nucleus is in the
color singlet state both before and after the interaction
\cite{yuri}. Averaging in the color space on the nucleon in figure
\fig{lo}A gives zero for this diagram since $\mbox{Tr} \, T^a = 0$.

Similar color arguments can rule out a number of diagrams with extra
gluons involved where only one gluon interacts with the
nucleon. Diagram in \fig{lo}B has two gluons interacting with the
nucleon and can not be proved to vanish by simple color
algebra. Adding the diagram with the gluon lines crossed to it and
performing color averaging one can see that the diagram is
proportional to
\ben
\delta ((p-l-l')^2) \, R^{a a'} \, \overline{\eta}^M_{a'-\mu} \, l_\mu \,
R^{b b'} \, \eta^M_{b'-\nu} \, l'_\nu \, \delta^{ab} \,
\left( \tilde{u} (p-l-l') \gamma_+ \frac{\gamma \cdot (p-l)}{(p-l)^2 + 
i \epsilon} \gamma_+ \tilde{u} (p) \, \right.
\een
\be\label{tgd1}
+ \, \left. \tilde{u} (p-l-l') \gamma_+ 
\frac{\gamma \cdot (p-l')}{(p-l')^2 + i \epsilon} \gamma_+ \tilde{u} (p) 
\right)
\ee
where $p$ is the momentum of the quark line in the nucleon below
having a large $p_+$ component. All the other light cone components of
the gluon's momenta in the diagrams we consider are much smaller than
that $l_+, l'_+ \, \ll \, p_+$ \cite{yuri}. Using the orthogonality of
the color rotation matrices $R^{a a'} \, R^{a b'} = \delta^{a'b'}$ and
evaluating the products of $\gamma$-matrices in the eikonal
approximation one can show that
\eq{tgd1} is proportional to
\be\label{deltas}
\overline{\eta}^M_{a-\mu} \, l_\mu \, \eta^M_{a-\nu} \, l'_\nu \, 
\delta (l_- + l'_-) \, \left( \frac{1}{l_- - i \epsilon} - 
\frac{1}{l_- + i \epsilon} \right) \, \sim \, l_- \, l'_- \, 
\delta (l_-) \, \delta (l'_-) \, = \, 0.
\ee
Thus we have shown that the contribution of the diagram in \fig{lo}B
to the action of the I-Nucleus system is also zero. 

We have to point out that due to the non-Abelian structure of the
effective instanton lagrangian of \eq{effl} there is another type of
an instanton vertex with two gluons fusing into an instanton at the
same point in coordinate space \cite{dpol}, which yields us with
another diagram which might contribute to the instanton--nucleus
interactions as shown in \fig{lo}C. However the vertex is
antisymmetric in the color indices of the gluons connecting to it. Due
to color averaging in the nucleon below the diagram in \fig{lo}C is
proportional to $f^{abc} \, \delta^{ab} \, = \, 0$ and does not
contribute to the action.

The reason why all the diagrams in \fig{lo} are zero could be
understood in terms of the effective instanton lagrangian result given
by \eq{effnf}. At the leading (classical) order the value of the
matrix elements of the local operators $G^{a}_{\mu\nu} (x_0)
G^{a}_{\mu\nu} (x_0)$ and $G^{a}_{\mu\nu} (x_0) \tilde{G}^{a}_{\mu\nu}
(x_0)$ in the nuclear wave function could be obtained by just
substituting the field strength of the classical non-Abelian
Weizs\"{a}cker-Williams field of a nucleus in it. As could be seen
from the exact expression for this field given in \cite{yuri} the only
non-vanishing components of its field strength tensor are
$G^{WWa}_{+\perp}$, so that $(G^{WWa}_{\mu\nu})^2 = G^{WWa}_{\mu\nu}
\tilde{G}^{WWa}_{\mu\nu} = 0$. Therefore from
\eq{effnf} it follows that there is no correction to the instanton
density if one calculates the matrix elements of $G^2$ and
$G\tilde{G}$ only at the classical level. However, there exists a
possibility that higher order in $\as$ corrections might yield a
non-zero value for these matrix elements. To check this we will now go
one step beyond and calculate what appears like a one loop correction
to this classical result.

\subsection{Next-to-Leading Order Diagrams}

\begin{figure}
\begin{center}
\epsfxsize=8cm
\leavevmode
\hbox{ \epsffile{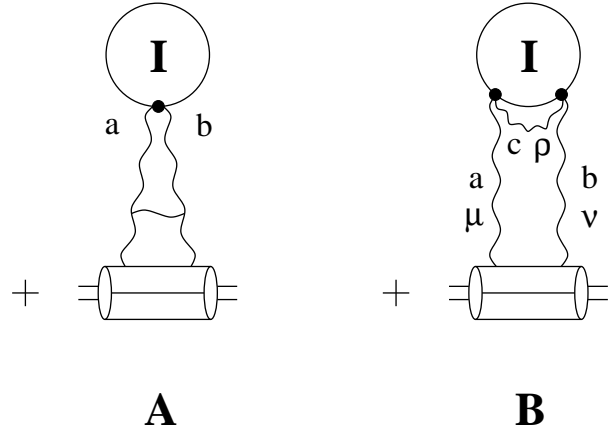}}
\end{center}
\caption{ Some of the two-loop graphs involving the instanton--two 
gluon vertex that vanish.}
\label{nlo1}
\end{figure}

In this subsection we are going to analyze two-loop diagrams
contributing to the action of the instanton-nucleus
configuration. These diagrams are of the order of $\rho^4$ in the
instanton size and involve only one interacting nucleon, that is only
one power of $\as^2 A^{1/3} \,
\lsim \, 1$ . First let us point out some two-loop diagrams which vanish 
or can be shown to be suppressed after a simple calculation. Those
include the diagrams where only one gluon interacts with the nucleus
and the diagrams with the vertices of the type introduced in
\fig{lo}C.  For instance the graph in \fig{nlo1}A vanishes for the same 
reason as the graph in \fig{lo}C: it is proportional to $f^{abc} \,
\delta^{ab} \, = \, 0$. All the diagrams generated by including all other 
possible connections of the extra gluon in \fig{nlo1}A to the gluon
and quark lines vanish for similar reasons of simple color algebra.
The case when the gluon connects to the instanton via the
instanton--two gluon vertex is a little different. To estimate the
diagram in \fig{nlo1}B we have to make use of the fact that the
nucleus is moving very fast in the light cone ``$+$'' direction. Then
the gluon propagators for the exchanged gluons give $g_{\mu-} \,
g_{\nu-}$ in covariant gauge. The contribution of the graph in
\fig{nlo1}B is therefore proportional to 
\be
g_{\mu-} \, g_{\nu-} \, \overline{\eta}^M_{d\mu\rho} \, R^{dd'} \,
f^{acd'} \, \eta^M_{e\nu\rho} \, R^{ee'} \, f^{bce'} \, \delta^{ab} \,
= \, - 3 \, N_c \, g_{--} \, = \, 0.
\ee

\begin{figure}
\begin{center}
\epsfxsize=12cm
\leavevmode
\hbox{ \epsffile{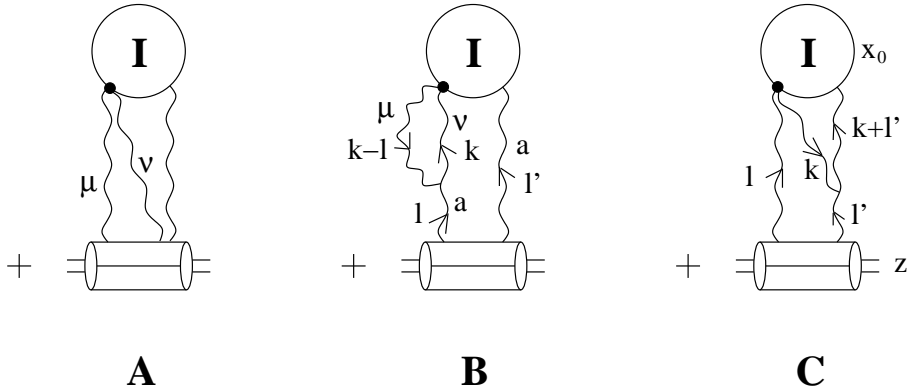}}
\end{center}
\caption{ A set of the two-loop graphs involving one instanton--double 
gluon vertex and one instanton--single gluon vertex that vanish.}
\label{nlo2}
\end{figure}

The gluon labeled $\rho$ in \fig{nlo1}B could also connect the
instanton--two gluon vertex to the other two gluon lines. These kind
of diagrams involve one instanton--single gluon vertex and one
instanton--double gluon vertex, as shown in \fig{nlo2}. The diagrams
in \fig{nlo2} can also be safely neglected for the following
reasons. The graph in \fig{nlo2}A is zero since in covariant gauge its
contribution is proportional to ${\overline \eta}^M_{a\mu\nu} \,
g_{\mu-} \, g_{\nu-} \, \sim \, g_{--} \, = \, 0$.

To prove that the contribution of the diagram in \fig{nlo2}B is zero
one has to add to it the diagram where the lines corresponding to $l$
and $l'$ gluons connect to the nucleon in the opposite order (crossed
diagram), which similarly to how in was shown in \eq{deltas} for the
graph of \fig{lo}B would give us two delta functions $\delta (l_-) \,
\delta(l'_-)$. Thus one can show that diagram of \fig{nlo2}B is 
proportional to
\ben
{\overline \eta}^M_{a\mu\nu} \, [ (k - 2 l)_\nu g_{\mu-} \, + \, (- 2
k + l)_- g_{\mu\nu} \, + \, (l + k)_\mu g_{\nu-} ] \, \eta^M_{a-\rho}
\, l'_\rho \, \delta (l_-) \, \delta(l'_-) \, = 
\een
\be 
= \,  - 3 {\overline \eta}^M_{a-\nu} \, l_\nu \, \eta^M_{a-\rho} \, 
l'_\rho \, \delta (l_-) \, \delta(l'_-) \, = \, 3 \, l_- \, l'_- \, 
\delta (l_-) \, \delta(l'_-) \, = \, 0.
\ee

Finally one can show that in the eikonal approximation employed here
the contribution of the graph in \fig{nlo2}C is independent of
$l_+$. Therefore the integral over $l_+$ in that diagram yields us
with \cite{yuri}
\be\label{noa}
\int_{-\infty}^{\infty} \frac{dl_+}{2 \pi} \, e^{- i l_+ (x_{0-} - z_-)} \, = \, 
\delta (x_{0-} - z_-)
\ee
where $x_0$ is the position of the instanton and $z_-$ is the (frozen)
light cone coordinate of the interacting quark in the
nucleon. \eq{noa} requires that the nucleon interacting with the
instanton at point $x_0$ should have the same light cone coordinate as
the instanton. But this would {\it not} allow us to sum over all
nucleons situated at different light cone coordinates at the given
impact parameter to obtain the enhancement of this diagram by the
powers of atomic number $A^{1/3}$ \cite{yuri,KM,claa}. In other words
the diagram of \fig{nlo2}C would not be resumming powers of the
parameter $\as^2 \, A^{1/3}$, it would have the factor of $\as^2$
without any $A^{1/3}$ enhancement. Thus it is zero at the leading
order in $A^{1/3}$.

\begin{figure}
\begin{center}
\epsfxsize=15cm
\leavevmode
\hbox{ \epsffile{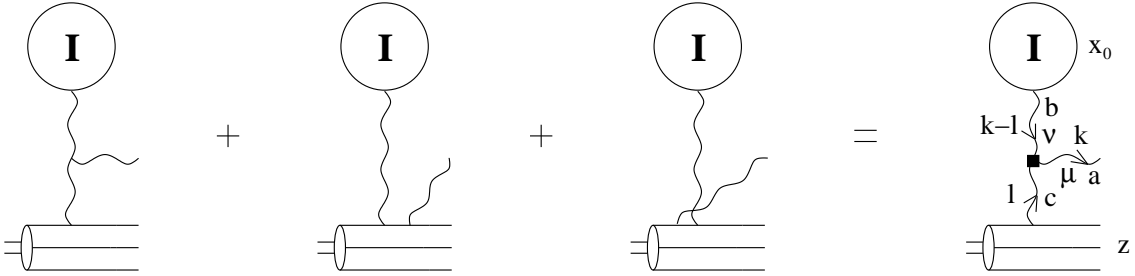}}
\end{center}
\caption{ Definition of the effective triple gluon vertex in 
$\partial_\mu A_\mu = 0$ covariant gauge.}
\label{effv}
\end{figure}

To start analyzing the diagrams with two instanton--single gluon
vertices let us define an effective triple gluon vertex as shown in
\fig{effv}. All the diagrams on the left of \fig{effv} are taken in
covariant gauge with the gluons' light cone momenta being much smaller
than the quarks light cone momentum. The graphs in \fig{effv}
correspond to the first non-trivial perturbative contribution to the
combined field of the instanton--nucleus configuration. An analogy
could be drawn with the Lipatov vertex being the first non-trivial
contribution to the field of two colliding nuclei
\cite{claa,KM,yuri1}. After a simple calculation one can see that the
lowest order field of the I--nucleus system is
\ben
A_\mu^a (k) \, = \, g f^{abc} (T^c) \, R^{bb'} \, 4 \pi^2 \rho^2 \int
\frac{d^4 l}{(2 \pi)^4} \, e^{-i l \cdot (x_0 - z)} \, (2 \pi) \delta
(l_-) \, \frac{(k-l)_\sigma}{(k-l)^2 + i \epsilon} \, \frac{1}{k^2 + i
\epsilon}
\een
\be\label{lof}
\times \left( - \frac{1}{{\underline l}^2} \, {\overline \eta}^M_{b'\nu\sigma} \,
[(k-2l)_\mu g_{\nu-} + (l+k)_\nu g_{\mu-} - 2 k_- g_{\mu\nu} ] +
{\overline \eta}^M_{b'-\sigma} \, g_{\mu-} \frac{1}{k_-} \right)
\ee
where we Fourier transform over the momentum $l$ flowing between the
nucleon at $z$ to the instanton at $x_0$. In \eq{lof} $(T^c)$ is the
$SU(3)$ matrix in the color space of the interacting nucleon and
${\underline l}$ denotes the transverse component of momentum $l$.

\begin{figure}
\begin{center}
\epsfxsize=7.5cm
\leavevmode
\hbox{ \epsffile{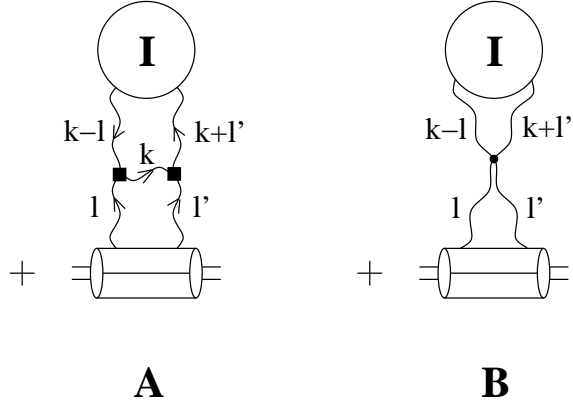}}
\end{center}
\caption{ Diagrams with two instanton--single gluon vertices that cancel 
each other at the leading order in $A$.}
\label{nlo}
\end{figure}

To calculate the contribution to the action at the order $\rho^4$ of
the field in \eq{lof} we have to consider the diagram depicted in
\fig{nlo}A. After averaging in the color space of the nucleon below 
\cite{yuri} the graph in \fig{nlo}A yields us with
\ben
i S_A \, = \, \frac{1}{2} \, \frac{g^2}{2} \, (4 \pi^2 \rho^2)^2 \,
\int \frac{d^4 k}{(2 \pi)^4} \, \frac{d^4 l}{(2 \pi)^4} \, 
\frac{d^4 l'}{(2 \pi)^4} \, e^{-i l \cdot (x_0 - z) - i l' \cdot (x_0 - z)} \, 
(2 \pi) \delta (l_-) \, (2 \pi) \delta (l'_-) \, \frac{-i}{k^2 + i
\epsilon}
\een
\ben
\times \frac{(k-l)_\sigma}{(k-l)^2 + i \epsilon} \, 
\frac{(k+l')_{\sigma'}}{(k+l')^2 - i \epsilon} \, 
\left( \frac{1}{{\underline l}^2} \, [- 
{\overline \eta}^M_{b-\sigma} \, (k-2l)_\mu - {\overline
\eta}^M_{b\nu\sigma} \, (l+k)_\nu g_{\mu-} + 
{\overline \eta}^M_{b\mu\sigma} \, 2 k_- ] + {\overline \eta}^M_{b
-\sigma} \, g_{\mu-} \frac{1}{k_-} \right)
\een
\be\label{acta1}
\times \left( \frac{1}{{\underline l'}^2} \, [- 
\eta^M_{b-\sigma'} \, (k+2l')_\mu - \eta^M_{b\nu'\sigma'} \, (k-l')_{\nu'} 
g_{\mu-} + \eta^M_{b\mu\sigma'} \, 2 k_- ] + \eta^M_{b
-\sigma'} \, g_{\mu-} \frac{1}{k_-} \right)
\ee
where the field on the right hand side of \fig{nlo}A is complex
conjugate to the field on the left hand side. The first factor of
$1/2$ in \eq{acta1} is the symmetry factor while the second factor of
$1/2$ arises after averaging over colors in the nucleon. Simplifying
the expression in \eq{acta1} by employing the algebra of 't Hooft
symbols \cite{th} we obtain
\ben
i S_A \, = \, 4 \pi^4 \rho^4 g^2 \int \frac{d^4 k}{(2 \pi)^4} \, 
\frac{d^4 l}{(2 \pi)^4} \, \frac{d^4 l'}{(2 \pi)^4} \, e^{-i l \cdot 
(x_0 - z) - i l' \cdot (x_0 - z)} \, (2 \pi) \delta (l_-) \, (2 \pi)
\delta (l'_-) \, \frac{-i}{k^2 + i
\epsilon}
\een
\be\label{acta2}
\times \frac{1}{(k-l)^2 + i \epsilon} \, \frac{1}{(k+l')^2 - i \epsilon} 
\, \frac{k_-^2}{{\underline l}^2 {\underline l}'^2 } \, [- 13 k^2 + 
14 k \cdot l - 14 k \cdot l' + 24 l \cdot l' - 5 {\underline l}^2 - 5
{\underline l}'^2].
\ee

To average over the nuclear wave function one has to average over the
positions of all the quarks in the nucleons and nucleons in the
nucleus and sum over all the nucleons and quarks in them
\cite{yuri}. This could be summarized by the following operation
\be\label{ave}
\sum_{\mbox{nucleons}} \, \, \sum_{\mbox{val. quarks}} \, 
\int \frac{d^2 z \, dz_-}{S_\perp 2 b_-}
\ee
which has to be applied to \eq{acta2}. For simplicity we consider a
cylindrical nucleus with cross sectional area $S_\perp$ and the
longitudinal extent of $2 b_-$ in the infinite momentum
frame. Applying \eq{ave} to \eq{acta2} yields for a quarkonium nucleus
of \cite{yuri,md}
\ben
i \left< S_A \right> \, = \, - i 4 \pi^4 \rho^4 g^2 \,
\frac{2 A}{S_\perp} \, \int
\frac{d^4 k}{(2 \pi)^4} \, \frac{d^2 l}{(2 \pi)^2} \, 
\frac{dl_+ dl'_+}{(2 \pi)^2} \, 
\frac{e^{- i l_+ x_{0-} - i l'_+ x_{0-}}}{k^2 [(k-l)^2 + i \epsilon] 
[(k+l')^2 - i \epsilon]} \, \frac{k_-^2}{({\underline l}^2)^2 } 
\een
\be\label{acta3}
\times [- 13 k^2 + 14 k \cdot l - 14 k \cdot l' + 14 {\underline l}^2 ].
\ee
Performing similar calculations for the diagram in \fig{nlo}B we
obtain
\be\label{actb1}
i \left< S_B \right> \, = \,  i 4 \pi^4 \rho^4 g^2 \,
\frac{2 A}{S_\perp} \, \int
\frac{d^4 k}{(2 \pi)^4} \, \frac{d^2 l}{(2 \pi)^2} \, 
\frac{dl_+ dl'_+}{(2 \pi)^2} \, 
\frac{e^{- i l_+ x_{0-} - i l'_+ x_{0-}}}{[(k-l)^2 + i \epsilon] 
[(k+l')^2 - i \epsilon]} \, \frac{k_-^2}{({\underline l}^2)^2 }.
\ee
Adding up the contributions of Eqs. (\ref{acta3}) and (\ref{actb1}) we
write
\ben
i \left< S_{A+B} \right> \, = \, - i 4 \pi^4 \rho^4 g^2 \,
\frac{2 A}{S_\perp} \, \int
\frac{d^4 k}{(2 \pi)^4} \, \frac{d^2 l}{(2 \pi)^2} \, 
\frac{dl_+ dl'_+}{(2 \pi)^2} \, 
\frac{e^{- i l_+ x_{0-} - i l'_+ x_{0-}}}{k^2 [(k-l)^2 + i \epsilon] 
[(k+l')^2 - i \epsilon]} \, \frac{k_-^2}{({\underline l}^2)^2 } 
\een
\be\label{actt}
\times (-7) \, [(k-l)^2 + (k+l')^2].
\ee
If one wants to perform the $l_+$ and $l'_+$ integrals in \eq{actt}
one has to pick up the poles given by the $(k-l)^2$ and $(k+l')^2$
denominators. However each of the terms in the square brackets at the
end of \eq{actt} cancels one of these denominators making the
expression in \eq{actt} zero for any non-zero $x_{0-}$. Thus the
diagrams in \fig{nlo} cancel each other and give zero at the leading
powers in $A$.

That way we have shown that even at one loop order the field of a
large nucleus does not affect the instanton distribution
\be\label{aeff}
n^A_{sat} (\rho) \, = \, n_0 (\rho).
\ee
The physical reason behind this result of our calculation is the
following. We have seen in Sect. IIIA that for a purely classical
field the only non-zero component of the field strength tensor is
$G^{WWa}_{+\perp}$ \cite{yuri,jklw}, which leads to $G^2 = G
\tilde{G} = 0$ and due to \eq{effnf} the field does not affect the
instanton size distribution. The diagrams we have analyzed above,
especially the graphs in \fig{nlo} correspond to including one rung
of the QCD evolution in energy. (We were interested in the real part
of the diagrams and did not get the factor of $\ln 1/x$ from them.)
Recently, in \cite{dip,JKLW} a picture of the evolution in $\ln 1/x$
has been developed which represents the quantum evolution as a series
of classical emissions. At each step of the evolution the existing
partons act as classical sources of color charge and emit a
gluon. The gluon then gets incorporated into the source and acts as a
color charge emitting gluons in the subsequent steps of the
evolution\footnote{This picture is based on the achieved long ago
understanding that the appearance of $\ln(1/x)$ for each emitted gluon
reflects the fact that this gluon lives much longer light cone time
than all gluons emitted after it but much shorter time than gluons
emitted before it \cite{grib}.}. This tells us that if we look at the
gluon field generated by any number of evolution steps it would
always look like a classical field emitted off some complicated
(evolved) source moving in the ``$+$'' direction and therefore has
only $G^{WWa}_{+\perp}$ non-zero component of the field strength
tensor. Thus from \eq{effnf} we derive that this field can not change
the instanton density. Multiple rescatterings, i.e., higher powers of
$\as^2 A^{1/3}$ would not help to obtain a non-zero effect since $G^2
= G \tilde{G} = 0$ for the full non-Abelian Weizs\"{a}cker-Williams
field of a large nucleus which includes the effects of all the
multiple rescatterings in it
\cite{yuri,jklw,KM}.

\begin{figure}
\begin{center}
\epsfxsize=2.1cm
\leavevmode
\hbox{ \epsffile{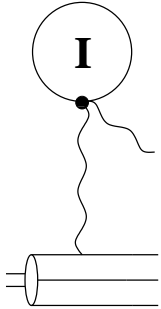}}
\end{center}
\caption{Last diagram which along with the graphs in \fig{effv} 
contributes to the topologically non-trivial classical field of a 
large nucleus at the lowest order. }
\label{field}
\end{figure}

Since the field of a single nucleus does not suppress instantons it
would be interesting to construct a solution of the Yang-Mills
equations of motion with the nuclear source carrying some non-zero
topological charge. The effective action of \eq{qcdi} allows us to do
that perturbatively. The lowest order field is just a direct sum of
the fields of the instanton and the nucleus.  At the lowest
non-trivial order in $\rho^2$ and $g$ almost all the field is given by
the diagrams contributing to the effective vertex of \fig{effv} and is
written down in \eq{lof}. Another diagram which contributes to the
field at this lowest order is shown in \fig{field} and involves and
instanton--double gluon vertex.

Adding the contribution of the diagram from \fig{field} to the
contribution of graphs from \fig{effv} we obtain the following
expression for the total field of the instanton-nucleus configuration
at the lowest order in momentum space
\ben
A_\mu^a (k) \, = \, g f^{abc} (T^c) \, R^{bb'} \, 4 \pi^2 \rho^2 \int
\frac{d^4 l}{(2 \pi)^4} \, e^{-i l \cdot (x_0 - z)} \, (2 \pi) \delta
(l_-) \, \frac{1}{k^2 + i
\epsilon} \, \left[ \frac{(k-l)_\sigma}{(k-l)^2 + i \epsilon} \right.
\een
\be\label{loft}
\times \left. \left( - \frac{1}{{\underline l}^2} \, 
{\overline \eta}^M_{b'\nu\sigma} \, [(k-2l)_\mu g_{\nu-} + (l+k)_\nu
g_{\mu-} - 2 k_- g_{\mu\nu} ] + {\overline \eta}^M_{b'-\sigma} \,
g_{\mu-} \frac{1}{k_-} \right) \, +
\, \frac{i}{{\underline l}^2} \, {\overline \eta}^M_{b'\mu-} \right]. 
\ee
To summarize the results of this section we once again point out that
the saturation effects fail to either enhance or suppress instantons
in a single hadron or nucleus (\eq{aeff}), which makes possible the
existence of a classical solution for the gluon field of the nucleus
carrying a non-zero topological charge, a lowest order expression for
which is given by \eq{loft} for the case of unit topological
charge. Now that we have demonstrated the formalism for calculating
the effects of a background classical field on the instanton
distribution we continue by considering the case of hadronic or
nuclear collisions.

\section{Hadronic and Nuclear Collisions}

It was shown in \cite{claa,md,kv,yuri1} that the gluon production in
the central rapidity region of a heavy ion collision at very high
energies is dominated by the classical gluonic field produced by two
colliding nuclei. The colliding nuclei could be visualized as
ensembles of point color charges moving without any deflection through
each other along the light cone \cite{claa}, similarly to the case of
a single nucleus in McLerran-Venugopalan model \cite{mv}. As the
nuclei pass through each other the color charges in each nucleus get
rotated in color space by the field of another nucleus, which leads to
a bremsstrahlung emission of gluons off these charges \cite{md}. The
gluons then subsequently multiply rescatter in the background fields
of both nuclei before being produced \cite{yuri1}.

The produced classical field is boost invariant and fills the whole
region between the nuclei moving apart after the collision
\cite{claa,md,kv,yuri1}. The field is produced at all the impact
parameters where the collision happens. Thus for central collisions
the classical field covers the whole nuclear cross sectional area in
the transverse direction $S_\perp$. In the longitudinal direction the
typical gluon production time is of the order of $2 k_+ /{\underline
k}^2$. Since for classically produced gluons $|{\underline k}| \sim
Q_s$ and $k_+ \sim Q_s$ \cite{yuri1,bmss,musat} the typical
longitudinal time is of the order of $1/Q_s$. This translates into the
typical thickness in the $z$-direction of the region in which the
classical fields are produced being of the order of $ l \sim
1/Q_s$. As was argues by Baier et al in \cite{bmss} even though the
classical field, as a leading order in $\as$ term, would still exist
at later times the thermalization effects would become important there
significantly modifying the distribution of produced gluons. In our
effective lagrangian approach we consider relatively small instantons
with sizes $\rho \ll l \sim 1/Q_s$ (see \eq{apprx}) which might easily
appear in the spatial region described above at times of the order
$\tau \sim 1/Q_s$. For such small instantons it is reasonable to
consider the problem of instantons in the classical field produced in
a nuclear collision before thermalization effects became important.

To estimate the effect of this classical gluon field produced in
mid-rapidity on instantons one has to calculate the field strength
tensor of this field and substitute it into \eq{effnf} averaging over
the wave functions of both colliding nuclei. An analytical expression
for the classical field exists only at the lowest non-trivial order
\cite{claa,md}. It is possible to calculate the multiplicity 
distribution of the produced particles analytically to all orders in
the background field without deriving an explicit expression for the
field \cite{yuri1}. Numerical results exist for the field and
multiplicity distribution \cite{kv}. We begin by calculating the field
strength tensor at the lowest order corresponding to the case when
$\as^2 A_1^{1/3} \lsim 1$ and $\as^2 A_2^{1/3} \lsim 1$, i.e., the
weak field case. $A_1$ and $A_2$ are the atomic numbers of the
colliding nuclei. In this case only one nucleon from each nucleus
contributes to the classical field.  We will later generalize our
result to the strong field case of $\as^2 A_1^{1/3} \sim 1$ and $\as^2
A_2^{1/3} \sim 1$.

\begin{figure}
\begin{center}
\epsfxsize=9cm
\leavevmode
\hbox{ \epsffile{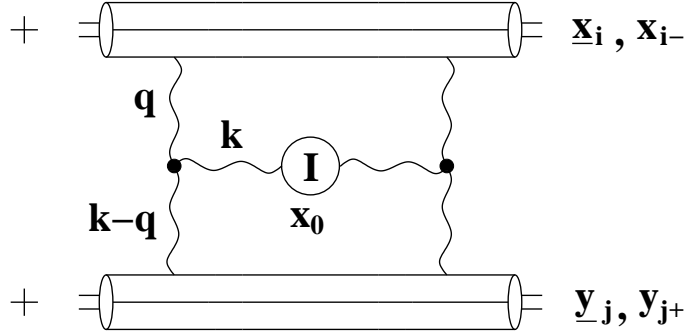}}
\end{center}
\caption{Instanton in the background gluon field of colliding nuclei, 
which is taken at the lowest non-trivial order in $g$. Thick dots
represent Lipatov vertices.}
\label{aa}
\end{figure}

The lowest order classical gluon field of two colliding nucleons $i$
and $j$ having coordinates $\underline{x}_i$, $x_{i-}$ and
$\underline{y}_j$, $y_{j+}$ correspondingly in covariant gauge
$\partial \cdot A = 0$ is given by
\cite{md}
\be\label{lofi}
A_{\mu}^{a} (x) = - i \int \frac{ d^4 k}{(2 \pi)^4}\, \frac{e^{- i k
\cdot x} }{k^2 + i \epsilon k_0 } \, \int d^2 q \, \frac{g^3}{ (2 \pi)^2} \, 
f^{abc}\,
 (T_i^b) \, (\tilde{T}_j^c)\, e^{i [k_+ x_{i-} + k_- y_{j+} -
 \underline{k} \cdot \underline{y}_j - \underline{q} \cdot
 (\underline{x}_i - \underline{y}_j)]} \, \frac{C_\mu (k,
 \underline{q})}{\underline{q}^2 (\underline{k} - \underline{q})^2} 
\ee
where $C_\mu (k, \underline{q})$ is the Lipatov vertex \cite{BFKL}
\be\label{lipa}
C_\mu (k, \underline{q}) \, = \, \left( \frac{ \underline{q}^2 }{k_- +
i \epsilon} - k_+ \, , \, - \frac{ (\underline{k} - \underline{q})^2
}{k_+ + i \epsilon} + k_- \, , \, 2\underline{q} - \underline{k}
\right).
\ee
In \eq{lipa} the four-vector $C_\mu$ is shown in terms of its
components in the $(+,-,\perp)$ form. $(T_i^b)$ and $(\tilde{T}_j^c)$
in \eq{lofi} are matrices in the color spaces of the colliding
nucleons \cite{yuri,md}. Regularization of the gluon propagator $k^2$
in \eq{lofi} corresponds to using retarded Green function and is used
to insure casuality of the classical field: it can be non-zero only in
the forward light cone \cite{md}.

Let us start by calculating the averaged value of the square of the
field strength tensor $\left< [G_{\mu\nu}^{a} (x_0)]^2 \right>$ to be
used in \eq{effnf} at the lowest order in $\as$. One can easily see
that at the lowest order in the coupling $g$ only the Abelian part of
$G_{\mu\nu}^{a}$ would contribute. We are interested only in the
$\left< [G_{\mu\nu}^{a} (x_0)]^2 \right>$ in the forward light cone in
the central rapidity region. Thus we do not need to include the
effects of the lowest order fields of each of the nuclei which give
non-zero $G_{\mu\nu}^{a}$ only on the light cone (at $x_+=0$ and/or
$x_-=0$). The diagram that we need to calculate is shown in \fig{aa}
and corresponds to the lowest order gluon field (order $g^3$)
interacting with the instanton at the lowest order in $\rho^2$ (order
$\rho^4$). Blobs in \fig{aa} denote Lipatov vertices.

The contribution of the diagram in \fig{aa} is
\be\label{g21}
\left< G_{\mu\nu}^{a} (x_0) G_{\mu\nu}^{a} (x_0) \right>_{LO} \, = \, \left<  
[\partial_\mu A_\nu^a (x_0) - \partial_\nu A_\mu^a (x_0)]^2 \right>
\ee
with $x_0$ the position of the instanton. Substituting the field of
\eq{lofi} into \eq{g21} and averaging over colors and positions of 
nucleons using \eq{ave} we obtain
\ben
\left< G_{\mu\nu}^{a} (x_0) G_{\mu\nu}^{a} (x_0) \right>_{LO} \, = \, 
\frac{g^6}{(2 \pi)^2} \, \frac{A_1 A_2}{S_{1\perp} S_{2\perp}} \, 2 \, C_F 
\, \int \ \frac{d^4 k}{(2 \pi)^4} \, \frac{dk'_+ dk'_-}{(2 \pi)^2} \, 
\frac{e^{- i k_+
x_{0-} - i k_- x_{0+} - i k'_+ x_{0-} - i k'_- x_{0+} } }{(k^2 + i
\epsilon k_0) (k'^2 + i \epsilon k'_0) } 
\een
\be\label{g22}
\times \int \frac{d^2 q}{[\underline{q}^2 (\underline{k} - \underline{q})^2]^2} 
\, [k_\mu \, C_\nu (k, \underline{q}) - k_\nu \, C_\mu (k, \underline{q})] \, 
[k'_\mu \, C_\nu (k', - \underline{q}) - k'_\nu \, C_\mu (k', -
\underline{q})]
\ee
where throughout \eq{g22} we imply that ${\underline k}' = -
{\underline k}$. In arriving at \eq{g22} we for simplicity assumed
that the colliding nuclei are cylinders in the $z$ direction with
cross sectional areas of $S_{1\perp}$ and $S_{2\perp}$
correspondingly. As was shown in \cite{md} in the integration over the
light cone components of $k$ and $k'$ only the poles in the
propagators $k^2$ and $k'^2$ contribute. If one picks up the poles in
the Lipatov vertices (see \eq{lipa}) the resulting contribution would
be non-zero only on the light cone \cite{md}. However here we are
interested in the central rapidity region only and that contribution
would not be important to us. Thus only the gluon propagator poles
would contribute in \eq{g22}. Anticipating this we may estimate the
value of the terms in the square brackets at the end of \eq{g22}
employing the $k^2 = 0$ and $k'^2 = 0$ conditions. After some lengthy
algebra one ends up with
\ben
[k_\mu \, C_\nu (k, \underline{q}) - k_\nu \, C_\mu (k, \underline{q})] \, 
[k'_\mu \, C_\nu (k', - \underline{q}) - k'_\nu \, C_\mu (k', -
\underline{q})] |_{k^2 = k'^2 =0} \, = 
\een
\be\label{lipsq}
= \, \frac{8 \, k \cdot k'}{\underline{k}^2} \, \left[ - 2 (\underline{k} \cdot 
\underline{q})^2 + \underline{k}^2 \underline{q}^2 + 2 (\underline{k} 
\cdot \underline{q}) \underline{q}^2 - (\underline{q}^2)^2 \right].
\ee
Substituting \eq{lipsq} into \eq{g22} yields
\ben
\left< G_{\mu\nu}^{a} (x_0) G_{\mu\nu}^{a} (x_0) \right>_{LO} \, = \, 
\frac{g^6}{(2 \pi)^2} \, \frac{A_1 A_2}{S_{1\perp} S_{2\perp}} \, 2 \, C_F 
\, \int \ \frac{d^2 k \, d^2 q}{(2 \pi)^4} \,  
\frac{8 [- \underline{q}^2 (\underline{k} - \underline{q})^2 + 
2 \underline{k}^2 \underline{q}^2 - 2 (\underline{k} \cdot 
\underline{q})^2]}{\underline{k}^2 [\underline{q}^2 
(\underline{k} - \underline{q})^2]^2} 
\een
\be\label{g23}
\times \,  \frac{dk_+ dk_- dk'_+ dk'_-}{(2 \pi)^2} \, \frac{e^{- i k_+
x_{0-} - i k_- x_{0+} - i k'_+ x_{0-} - i k'_- x_{0+} } }{(k^2 + i
\epsilon k_0) (k'^2 + i \epsilon k'_0) } \, k \cdot k' .
\ee
One can readily check that
\be\label{j1}
\int d^2 q \, \frac{[- \underline{q}^2 (\underline{k} - \underline{q})^2 + 
2 \underline{k}^2 \underline{q}^2 - 2 (\underline{k} \cdot
\underline{q})^2]}{[\underline{q}^2 
(\underline{k} - \underline{q})^2]^2} \, = \, -
\frac{\pi}{\underline{k}^2}.
\ee
To obtain \eq{j1} it is easier to first integrate over the angles
between $\underline{k}$ and $\underline{q}$ after which the
integration over $|\underline{q}|$ becomes trivial. Employing \eq{j1}
in \eq{g23} we arrive at
\ben
\left< G_{\mu\nu}^{a} (x_0) G_{\mu\nu}^{a} (x_0) \right>_{LO} \, = \, 
- \frac{g^6}{(2 \pi)^2} \, \frac{A_1 A_2}{S_{1\perp} S_{2\perp}} \, 2 \, C_F 
\, 8 \pi \, \int \ \frac{d^2 k}{(2 \pi)^4 (\underline{k}^2)^2} 
\een
\be\label{g24}
\times \, \frac{dk_+ dk_- dk'_+ dk'_-}{(2 \pi)^2} \, \frac{e^{- i k_+
x_{0-} - i k_- x_{0+} - i k'_+ x_{0-} - i k'_- x_{0+} } }{(k^2 + i
\epsilon k_0) (k'^2 + i \epsilon k'_0) } \, k \cdot k' .
\ee
Evaluating the integral over the light cone components of $k$ and $k'$
in \eq{g24} we note that the integral over $k_\perp$ in \eq{g24} is
dominated by small transverse momenta $k_\perp$. Taking the $k_\perp
\rightarrow 0$ limit of the longitudinal integral we obtain
\be\label{t1}
\int \frac{dk_+ dk_- dk'_+ dk'_-}{(2 \pi)^2} \, \frac{e^{- i k_+
x_{0-} - i k_- x_{0+} - i k'_+ x_{0-} - i k'_- x_{0+} } }{(k^2 + i
\epsilon k_0) (k'^2 + i \epsilon k'_0) } \, k \cdot k' \, \approx \, 
\frac{1}{\tau_0^2},
\ee
where $\tau_0 = \sqrt{2 x_{0+} x_{0-}}$ is the proper time. Here we
made use of the fact that we are interested in instantons in the
forward light cone $x_{0+} > 0$ and $x_{0-} > 0$. The approximation
used in obtaining \eq{t1} is equivalent to taking the limit of early
proper time, which more formally means $\tau_0 Q_s \ll 1$. This
approximation is not crucial for our approach and is made only to
simplify the calculations.

Employing \eq{t1} in \eq{g24} yields
\be\label{g25}
\left< G_{\mu\nu}^{a} (x_0) G_{\mu\nu}^{a} (x_0) \right>_{LO} \, = \, 
- \as^3 \, \frac{A_1 A_2}{S_{1\perp} S_{2\perp}}
\, C_F \, \frac{16}{\pi^2} \, \frac{1}{\tau_0^2} \, \int \, 
\frac{d^2 k}{(\underline{k}^2)^2}. 
\ee

Before we discuss the issue of regularization of the divergent
integral over transverse momentum in \eq{g25} let us first evaluate
the other correlator in the exponent of \eq{effnf}. The lowest order
field of \eq{lofi} yields after averaging over nucleons similarly to
\eq{g22}
\ben
\left< G_{\mu\nu}^{a} (x_0) \tilde{G}_{\mu\nu}^{a} (x_0) \right>_{LO} \, = \, 
\frac{g^6}{(2 \pi)^2} \, \frac{A_1 A_2}{S_{1\perp} S_{2\perp}} \, 2 \, C_F 
\, \int \ \frac{d^4 k}{(2 \pi)^4} \, \frac{dk'_+ dk'_-}{(2 \pi)^2} \, 
\frac{e^{- i k_+ x_{0-} - i k_- x_{0+} - i k'_+ x_{0-} - i k'_- x_{0+} } }
{(k^2 + i \epsilon k_0) (k'^2 + i \epsilon k'_0) } 
\een
\be\label{ggd1}
\times \int \frac{d^2 q}{[\underline{q}^2 (\underline{k} - \underline{q})^2]^2} 
\, 2 \, \epsilon_{\mu\nu\rho\sigma} \, k_\mu \, C_\nu (k, \underline{q}) \, 
k'_\rho \, C_\sigma (k', - \underline{q}) 
\ee
where again ${\underline k}' = - {\underline k}$. After some simple
algebra one obtains
\be\label{ggd2}
\epsilon_{\mu\nu\rho\sigma} \, k_\mu \, C_\nu (k, \underline{q}) \, 
k'_\rho \, C_\sigma (k', - \underline{q}) |_{k^2 = k'^2 = 0} \, = \, 8
\, \epsilon_{\mu\nu} \, k^\perp_\mu \, q^\perp_\nu \, \underline{q} \cdot 
(\underline{k} - \underline{q}),
\ee
where we have again anticipated that the integration over the
longitudinal momenta has to pick up the poles of the gluon propagators
giving $k^2 = k'^2 = 0$ similarly to the way we used it in obtaining
\eq{lipsq}. \eq{ggd2} together with \eq{ggd1} gives
\be\label{ggd3}
\left< G_{\mu\nu}^{a} (x_0) \tilde{G}_{\mu\nu}^{a} (x_0) \right>_{LO} \, 
\sim \, \int \frac{d^2 q}{[\underline{q}^2 (\underline{k} - \underline{q})^2]^2} 
\, \epsilon_{\mu\nu} \, k^\perp_\mu \, q^\perp_\nu \, \underline{q} \cdot 
(\underline{k} - \underline{q}) \, = \, 0.
\ee
The fact that the integral over $q_\perp$ in \eq{ggd3} is zero could
be seen by changing variables $\underline{q} \rightarrow \underline{k}
- \underline{q}$ which would demonstrate that the integral is equal to
its negative. We conclude therefore that at this lowest non-trivial
order in the coupling constant
\be\label{ggd4}
\left< G_{\mu\nu}^{a} (x_0) \tilde{G}_{\mu\nu}^{a} (x_0) \right>_{LO} \, = \, 0. \label{topc}
\ee
We think that (\ref{topc}) is a general result required by the
symmetry of the problem. Indeed, since the topological charge
(\ref{topc}) determines the net helicity of the system, and {\it on
the average} the net helicity generated in the collision is equal to
zero, the expectation value of the topological charge should
vanish. This does not, however, imply a zero dispersion in the
topological charge distribution. The event-by-event fluctuations of
topological charge can exist, and would induce parity--odd
correlations in the multiparticle production \cite{Pbubble}.

The only non-zero correlator for the classical gluon field produced
in a hadronic or nuclear collision is thus $\left< G_{\mu\nu}^{a}
(x_0) G_{\mu\nu}^{a} (x_0) \right>$. The correlator in \eq{g25} has
an infrared-divergent integral in it. The divergence is similar to
the infrared singularity present in multiplicity distribution of the
produced gluons given by the lowest order perturbative diagram in
\fig{aa} (without the instanton). As was argued in
\cite{claa,md,kv,yuri1,musat} multiple rescatterings would regularize
the integral by effectively inserting an infrared cutoff in the
$k_\perp$ integral in \eq{g25} which would be proportional to the
saturation scale $Q_s$. Unfortunately we do not know exactly the
coefficient of proportionality between this effective cutoff and
$Q_s$. At the same time this coefficient would be very important for
evaluating the integral in \eq{g25}.

There are several ways to regularize the integral in
\eq{g25}. Saturation effects which regularize the integral in the
infrared could be included in the classical (multiple rescattering)
approximation \cite{mv,yuri,jklw,KM,yuri1,musat} or in a more
realistic way incorporating the effects of quantum evolution in energy
\cite{glrmq,LR,agl,yurieq,dip,bal,JKLW,cons,lt,braun,lub,IM}. The latter 
way involves resummation of leading logarithms of energy (i.e. powers
of $\as \ln 1/x$) which arise from developing additional soft partons
in the nuclear wave functions that will be produced in the
collision. However, as was argued in \cite{JKLW,KLM} the effect of
this quantum evolution is only to produce sources of color charge off
which the classical field would be emitted. This statement was
quantified in \cite{JKLW,cons,IM} for a single nucleus. Unfortunately
similar analysis has not been carried out for the gluon production in
nuclear collisions. Thus we will restrict ourselves to the case of
classical field emitted off the valence quarks in the nucleons of the
colliding nuclei \cite{claa,md}. The quantum evolution will be
suppressed implying that $\as \ln 1/x \, \lsim \, 1$ and no extra
gluons are produced. Of course generalization of \eq{g25} to include
the full effect of saturation strictly speaking does not reduce to
the problem of classical gluon production of \cite{claa,md} due to the
non-Abelian nature of $G_{\mu\nu}^a$ which leads to appearance of
three-$A_\mu$ and four-$A_\mu$ correlators in the full correlator
$\left< G_{\mu\nu}^{a} (x_0) \tilde{G}_{\mu\nu}^{a} (x_0)
\right>$. Thus the calculations presented below should be understood 
as an estimate of what the exact answer should be in the strong field
case.

We are going to conjecture the following procedure of generalizing the
result of \eq{g25} to the strong field case of $\as^2 A_1^{1/3} \sim
1$ and $\as^2 A_2^{1/3} \sim 1$. Let us first note that at the same
order in the coupling constant as was employed in \eq{g25} the
multiplicity of gluons produced in a collision of two quarkonium
nuclei is given by \cite{md}
\be\label{multlo}
\frac{dN}{d^2 b \, dy} \, = \, \as^3 \, \frac{A_1 A_2}{S_{1\perp} S_{2\perp}} \,
C_F \, \frac{32}{\pi} \, \int \, \frac{d^2 k}{(\underline{k}^2)^2} \,
\ln \frac{k_\perp}{\Lambda}
\ee
with $\Lambda$ some infrared cutoff, $b$ the impact parameter and $y$
rapidity of the produced gluon. As one can see the expression in
\eq{multlo} has an infrared singularity similar to the one in \eq{g25}. 
Assuming that both singularities get regulated in a similar fashion we
write with logarithmic accuracy
\be\label{ggn}
\left< G_{\mu\nu}^{a} (x_0) G_{\mu\nu}^{a} (x_0) \right> \, \approx \, 
- \frac{1}{2 \, \pi \, \tau_0^2} \,  \frac{dN}{d^2 b \, dy}.
\ee
We thus suppose that \eq{ggn} holds at all higher orders after
inclusion of multiple rescatterings and is independent of whether we
take a quarkonium model of the nuclei or realistic heavy ions.  It has
been conjectured in \cite{musat} that the multiplicity of the produced
gluons including all higher order multiple rescattering effects is
proportional to the multiplicity of gluons in the nuclear wave
function before the collision so that for the case of two identical
colliding cylindrical nuclei
\be\label{fmult}
\frac{dN}{d^2 b \, dy} \, = \, c \, \frac{C_F Q_s^2}{\as 2 \pi^2}.
\ee
\eq{fmult} has been written in \cite{musat} with logarithmic accuracy.
The proportionality coefficient in \eq{fmult} has been estimated
numerically to be $c = 1.29 \pm 0.09$ \cite{kv}. An analytical
calculation of \cite{yuri1} gives $c \approx 2 \ln 2 \approx 1.39$ and
RHIC data seems to suggest that $c = 1.23 \pm 0.20$
\cite{KN}. Inserting \eq{fmult} into \eq{ggn} yields
\be\label{fgg}
\left< G_{\mu\nu}^{a} (x_0) G_{\mu\nu}^{a} (x_0) \right> \, \approx \, - c \, 
 \frac{C_F Q_s^2}{4 \, \pi^3 \, \as \, \tau_0^2},
\ee
which, after being substituted in \eq{effnf} gives the distribution of
instantons
\be\label{naa}
n_{sat}^{AA} (\rho) \, = \, n_0 (\rho) \, \exp \left( - \frac{c \,
\rho^4 Q_s^4 } { 8 \, \as^2 \, N_c \, (Q_s \tau_0)^2 } \right).
\ee
\eq{naa} is our main result. It shows that large size instantons are 
suppressed by the strong classical fields generated in the nuclear
collision. As was argued in \cite{bmss} classical fields are a good
description of the produced gluonic medium only at the early times
$\tau_0 Q_s \lsim 1$. Thus for $\tau_0 \sim 1/Q_s$ the suppression
starts parametrically when $\rho^2 Q_s^2 \sim \as \ll 1$, that is our
main assumption of small instantons stated in \eq{apprx} is still
valid.

Three remarks are in order here. First let us note that the instanton
suppression of \eq{naa} results from \eq{effnf} applied to the case of
nuclear collisions. \eq{effnf} was originally derived for the
instanton distribution in the background of the color field generated
by QCD vacuum fluctuations as considered in
\cite{cdg,svz}. By definition of the physical vacuum the energy
density in it is negative with respect to perturbative vacuum. Thus
for the field of vacuum fluctuations $\left< \theta_\mu^\mu \right> <
0$. Recalling that for QCD in the chiral limit of massless quarks
\be\label{emt}
\left< \theta_\mu^\mu \right> \, = \, - \frac{b g^2}{32 \pi^2} \, 
\left< G_{\mu\nu}^{a2} \right>
\ee
we conclude that for vacuum fluctuations $\left< G_{\mu\nu}^{a2}
\right> > 0$. Therefore \eq{effnf} with the background field of vacuum
fluctuations having positive $\left< G_{\mu\nu}^{a2} \right>$ led the
authors of \cite{cdg,svz} to conclude that the large size instantons
are {\it enhanced} in QCD vacuum. The phenomenon is also referred to
as instanton melting. The situation is very different for nuclear
collisions. The energy density of the gluon matter produced in a heavy
ion collision is positive with the trace of the energy-momentum tensor
$\left< \theta_\mu^\mu \right> > 0$. With the help of
\eq{emt} we conclude that produced gluon matter should have 
$\left< G_{\mu\nu}^{a2} \right> < 0$, which agrees with \eq{fgg}. Thus
\eq{effnf} with negative $\left< G_{\mu\nu}^{a2} \right>$ leads to 
{\it suppression} of large size instantons in nuclear collisions as
appears in \eq{naa}.

Secondly, it may appear that the instanton suppression of \eq{naa}
contradicts the model of soft pomerons in hadronic collisions proposed
by the authors in \cite{kkl}. There each rung of the soft pomeron
ladder was modeled by the instanton-induced transitions of two
t-channel gluons into any number of s-channel gluons. If the
instantons were suppressed so would be the effects described in
\cite{kkl}. However the soft pomeron describes the total cross
sections in the proton-proton scattering in the kinematical range of
the Tevatron. There, even though the energies are high the colliding
particles are protons ($A=1$), not nuclei and unitarization effects
are still weak if present at all. The corresponding saturation scale
is presumably quite small, possibly being of the order of
$\Lambda_{QCD}$. Our suppression of \eq{naa} would be
indistinguishable from the ``usual'' suppression of large size
instantons in the instanton gas models of QCD vacuum \cite{ss} and
therefore would not significantly alter the results of
\cite{kkl}. Alternatively one could say that the gluon fields generated 
in proton-proton collisions at Tevatron energies are not as strong as
the fields produced in nuclear collisions at similar energies and thus
can not introduce a strong suppression of instantons.

When the energy of the proton-proton collisions becomes extremely
large saturation and unitarization effects begin to take place. As was
argued in \cite{glrmq,LR,mv,jklw,musat} at sufficiently high energies
the saturation scale even in proton-proton collisions would become
large, much larger than $\Lambda_{QCD}$, making most of the partons in
the protons' wave functions perturbative. At these high energies the
hard (BFKL) pomeron would be unitarized through multiple pomeron
exchanges \cite{agl,yurieq,dip,bal,JKLW}, which in fact would generate
the large saturation scale $Q_s$. At the same time unitarization of
the soft pomeron is not quite understood. There is a possibility that
the soft pomeron unitarizes through multiple pomeron exchanges,
similar to the hard pomeron. Here we would like to outline another
possibility, inspired by the soft pomeron model of \cite{kkl} and by
the result of \eq{naa}. The instanton-induced interactions leading to
the soft pomeron behavior of the cross section would be suppressed in
pp at high energies due to instanton suppression of \eq{naa}. Thus the
soft pomeron of \cite{kkl} could be eliminated by the large
perturbative saturation scale at extremely high energies and the cross
sections would be dominated by hard perturbative interactions which
are unitary.

\begin{figure}
\begin{center}
\epsfxsize=12cm
\leavevmode
\hbox{ \epsffile{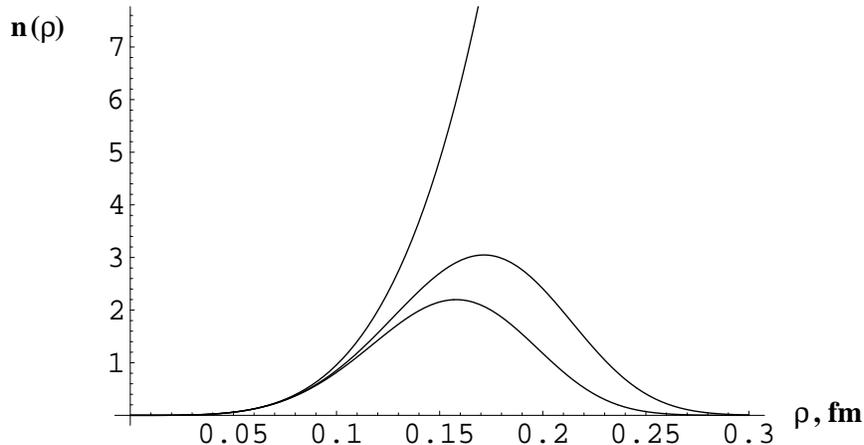}}
\end{center}
\caption{Distributions of instanton sizes in vacuum for QCD with 
three light flavors (upper curve) versus the distribution of instanton
sizes in the saturation environment produced by a collision of two
identical nuclei for $c=1$ (middle curve) and $c = 2 \ln 2$ (lower
curve) with $Q_s^2 \, = \, 2 \, \mbox{GeV}^2$ as estimated for RHIC at
$\sqrt{s} \, = \, 130 \, \mbox{AGeV}$ in \protect\cite{KN}.}
\label{nsat}
\end{figure}

Third we have to note that the suppression of \eq{naa} is different
from the instanton suppressions at finite $T$ and $\mu$. There the
power of the exponent is quadratic in $\rho$ (see Eqs. (\ref{fint})
and (\ref{finm})) while in our case it is quadric in $\rho$. The
difference is probably due to different physical mechanisms leading to
instanton suppression. At high temperature or density the interactions
of the gluons and quarks with each other generate a screening mass for
gluons $m_D$. The gluoelectric fields are therefore screened at
distances roughly of the order of $1/m_D$, which leads to instanton
suppression as discussed in the Introduction. In our case the gluon
screening on the transverse scales of order $1/Q_s$ is due to multiple
rescatterings in the background classical field generated not by the
other gluons but by the nuclei themselves. Therefore our expression in
\eq{naa} is different from the high temperature and high density
expressions in Eqs. (\ref{fint}) and (\ref{finm}).

In \fig{nsat} we have plotted the perturbative distribution of
instantons in vacuum $n_0 (\rho)$ of \eq{isize} including three light
flavors in the beta function $b = 9$ (upper curve) together with the
suppressed distributions of instantons in heavy ion collisions given
by \eq{naa} for two different values of the ``liberation
coefficient'' $c$: the middle curve corresponds to $c = 1$ and the
lower curve corresponds to $c = 2 \ln 2$. Saturation scale for the
identical cylindrical nuclei was taken to be $Q_s^2 \, = \, 2 \,
\mbox{GeV}^2$ based on recent RHIC data \cite{pho,KN}. The strong
coupling constant at this scale was approximated by $\as (Q_s)
\approx 0.3$. The suppression curves are plotted for relatively late
proper time $\tau_0 = 1/Q_s$. Instanton suppression is even stronger
for earlier times.

The integral of $n_0 (\rho)$ over all $\rho$ gives the total
instanton density at a given space-time point. Of course we can not
integrate the perturbative size distribution of \eq{isize} over all
values of $\rho$ since it diverges in the infrared. Instead we will
employ the fit to lattice data for the instanton distribution in pure
gluodynamics from \cite{rslat}. The density of instantons given by
this fit is $10.8 \, \mbox{fm}^{-4}$. Integration of \eq{naa} with
$n_0 (\rho)$ given by the fit of \cite{rslat} and with $Q_s^2 = 2 \,
\mbox{GeV}^2, \, c = 2 \ln 2$ yields us with the instanton density at
$\tau_0 = 1/Q_s$ of approximately $0.007 \, \mbox{fm}^{-4}$. The ratio
of the two numbers gives the overall suppression of instantons density
in the central rapidity region at RHIC to be $0.0006$. This estimate
is made for the case of pure gluodynamics and the number will of
course change once the effects of quarks are included. Nevertheless we
still expect to have approximately three orders of magnitude
suppression of instantons.  Since the saturation scale increases with
center of mass energy \cite{glrmq,yurieq,bal,lt} we expect the
suppression to get stronger for $\sqrt{s} \, = \, 200 \, \mbox{AGeV}$
at RHIC and for LHC energies. Numerical estimates of
\cite{braun,lub} give the increase of $Q_s(x)$ for gold nuclei
approximately by a factor of two for the LHC energy
\cite{braun,lub}. Therefore the expected suppression will be 
considerably stronger.

We hope that our results open the possibility of a systematic
theoretical investigation of topological effects in high energy
nuclear collisions and introduce a different angle of looking at the
problem of the interface between ``hard" (perturbative) and ``soft"
(non-perturbative) interactions.

\section*{Acknowledgments}

We would like to thank Ian Balitsky, Dmitri Diakonov, Asher Gotsman,
Uri Maor, Larry McLerran, John Negele, Victor Petrov, Rob Pisarski,
Maxim Polyakov, Edward Shuryak and Larry Yaffe for many informative
and stimulating discussions.  The research of D.K. is supported by the
U.S. Department of Energy under contract No. DE-AC02-98CH10886. The
work of Yu.K. was supported in part by the U.S. Department of Energy
under Grant No. DE-FG03-97ER41014. The research of Yu. K. and
E. L. was sponsored in part by the BSF grant $\#$ 9800276 and by
Israeli Science Foundation, founded by the Israeli Academy of Science
and Humanities.

\end{document}